\documentclass[aps,pre,twocolumn,showpacs,superscriptaddress,groupedaddress]{revtex4}
\usepackage{amsmath}
\usepackage{graphicx}
\usepackage{natbib}
\usepackage[toc,page]{appendix}
\usepackage[usenames]{color}
\usepackage[T1]{fontenc}
\usepackage{textcomp}
\usepackage{rotating}
\usepackage{etoolbox}
\begin{document}


\title{Replicator dynamics with diffusion on multiplex networks.}
\author{R.~J.~Requejo}
\author{A.D\'iaz-Guilera}
\affiliation{Departament de F\'isica Fonamental, Universitat de Barcelona, Mart\'i i Franques 1, 08028 Barcelona, Spain.}
\email {rrequejo@ffn.ub.edu}


\begin{abstract}
In this study we present an extension of the dynamics of diffusion in multiplex graphs which makes the equations compatible with the replicator equation with mutations. We derive an exact formula for the diffusion term, which shows that, while diffusion is linear for numbers of agents, it is necessary to account for non-linear terms when working with fractions of individuals. We also derive the transition probabilities that give rise to such macroscopic behavior, completing the bottom-up description. Finally, it is shown that the usual assumption of constant population sizes induces a hidden selective pressure due to the diffusive dynamics, which favors the increase of fast diffusing strategies.  
\end{abstract}


\maketitle

\section{Introduction}
During the last decade agent based modeling has increased its importance as a powerful tool to model situations in which the complexity of the interactions of many-agent systems makes it impossible, or at least very difficult, to make analytic predictions of the dynamical behavior of the system \cite{bonabeau:2002,grimm:2005}. Furthermore, the agent based models complement the analytic approach allowing for an exploration of the coarse-grained dynamics and the connection between micro-scale and macro-scale behavior \cite{helbing:1996, helbing:1998, traulsen:2005, traulsen:2006a}. Such agent based modeling is of special importance in evolutionary game theoretical studies \cite{maynard-smith:1982,sigmund:2010} which aim to capture the intricacies of biological, social and economical systems, where the non-linearity and feed-backs of the systems cannot be easily foreseen \cite{requejo:2011, requejo:2012, vilone:2011, requejo:2012c}. In the middle of such framework \cite{page:2002}, and connected with the micro-evolutionary dynamics \cite{helbing:1996,helbing:1998,traulsen:2005, traulsen:2006a}, stands the replicator equation \cite{taylor:1978, hofbauer:1979}
\begin{equation}\label{eq:repeq}
(\dot{x}^\alpha)_\textnormal{rep}=x^\alpha(f^\alpha-\bar{f})
\end{equation}

The replicator equation, which was introduced shortly after the foundation of the evolutionary game theoretical framework \cite{maynard-smith:1973,maynard-smith:1974,maynard-smith:1982}, has been extensively used during the last decades to model the evolution of the fractions $x^\alpha=n^\alpha/N$ of traits of type $\alpha$, $\alpha=1,...,L$ in large well-mixed populations with frequency dependent selection, i.e. when the fitness $f^\alpha$ (reproductive potential or capacity) of the agents traits and the mean population fitness $\bar{f}=\sum_\alpha f^\alpha x^\alpha$ depend on the population composition. Such traits may represent different phenotypes and genotypes in biological settings, or different behavioral strategies in a cultural evolutionary framework. 

In addition to classical biological applications related to the evolution of gene frequencies and phenotypic traits, 
the replicator equation and agent based simulations using microscopic updating rules that lead to it have been used to study how evolution, with a stress on the evolution of cooperation, is affected by physical entities. These studies include the interplay between different temporal scales of interaction and selection \cite{roca:2006, roca:2009}, the effect of network structures, both spatial lattices \cite{nowak:1992, roca:2009a, roca:2009} and scale-free networks \cite{santos:2005, santos:2008}, and the effect of linking fitness and resource availability \cite{requejo:2011,requejo:2012,requejo:2012b,requejo:2012c}, showing a self-organized feedback similar to homeostatic regulation \cite{lovelock:1974}. Furthermore, the richness of dynamical portraits \cite{bomze:1983, bomze:1995} has also been shown in minimalist scenarios, as exemplified by the introduction of loner \cite{hauert:2002a, hauert:2002b} and joker \cite{arenas:2011, requejo:2012} strategies, which allow for several phase transitions and complex dynamical behavior \cite{requejo:2012, hauert:2008, wakano:2009}, or the use of the discrete time version of the replicator equation for two strategies, which has been proven to show periodic and chaotic behavior \cite{vilone:2011}.

The replicator equation is on the core of the framework of evolution \cite{page:2002}, linked to the quasispecies equation \cite{page:2002}, the game dynamical equation \cite{helbing:1996, helbing:1998}, adaptive dynamics \cite{page:2002}, Lotka-Volterra dynamics \cite{hofbauer:1981,hofbauer:1998} and the Price equation \cite{price:1970, price:1972, price:1972a, traulsen:2010}. The extension of the replicator dynamics to regular networks has been developed in a situation where each node is an agent \cite{ohtsuki:2006b}, finding that for weak selection it corresponds to a transformation of the payoff matrix. Furthermore, an extension of the replicator dynamics has been developed to represent a two dimensional world where the agents diffuse \cite{wakano:2009}, which shows the appearance of Turing patterns and other complex behavior. However, many real world situations cannot be modeled as a simple network or a two dimensional space, and need the introduction of several kinds of links to represent different properties of the system \cite{kivela:2014}, as in the air transportation networks, where each airline represents a different network \cite{cardillo:2013}. 

Let us focus on a cultural evolutionary framework along the rest of the paper. In such a framework the individual traits being selected are behavioral traits, called strategies. The main aim of this paper is to develop the mathematical tools that allow to model diffusion of strategies in multiplex networks in a compatible way with the selection dynamics described by the replicator equation (with or without mutations), irrespective of the microscopic dynamics that give rise to the replicator equation (see appendix~\ref{app:a}). Let us provide a simple example to illustrate the situation: imagine several cities which are connected by bus, train and plane, each kind of connection with a different network structure. Individuals in each city are interacting between them, both directly or indirectly: they have different jobs and different incomes, which determines, at least partially, their choices on how to travel, and they also have information about the transportation ways. Based on such interactions and information, they may decide to travel in one or another transport; such choice determines the individual strategy. As the different transportation ways determine different networks structures, as well as the travel speeds are different, a multi-layer network is necessary to represent the full transportation system; hence, as a first approach to such simplified situation, we need to extend the replicator dynamics (describing strategy changes) to diffusive individuals in this kind of multi-layer.

We may think that introducing diffusion on multiplex models of evolution is a straightforward task, as the diffusive process has extensively been studied on such networks \cite{gomez:2013, sole-ribalta:2013}. However, it poses a challenge: as the replicator dynamics describe the evolution of fractions of individuals, the diffusive models need to be rewritten in a compatible way, accounting for the constraint on the addition of the fractions to one, as well as for the conservation of the number of agents. We develop such extension in this paper, showing that working with fractions introduces dependencies which are not present for the diffusion of the numbers of agents. These new dependencies can only be taken into account introducing a non-linear term, which has to be added to the linear one, in order to represent the general dynamics of diffusion of fractions of individuals in the multiplex. Furthermore, we discuss some situations in which the linear scenario can be recovered, as when population sizes or population size ratios between sites are constant, and show that in such situations hidden selective pressures act on the system, even when they do not appear explicitly on the equations.

The paper is outlined as follows: We start in section~\ref{sec:fractions} by defining multiplex networks and showing the problem to overcome when working with fractions; after that, in section~\ref{sec:diff} we derive the diffusion term compatible with the replicator dynamics and infer the transition probabilities that give rise to it; then, in section~\ref{sec:linear} we discuss some situations in which the linear dynamics are recovered (and the analytic calculations simplified), and show that some of this situations carry attached the appearance of hidden selective pressures; finally, in section~\ref{sec:conclusions} we discuss the results.

\section{Diffusion in the multiplex: The problem of working with fractions.}
\label{sec:fractions}

\begin{figure}
\includegraphics[width=\linewidth]{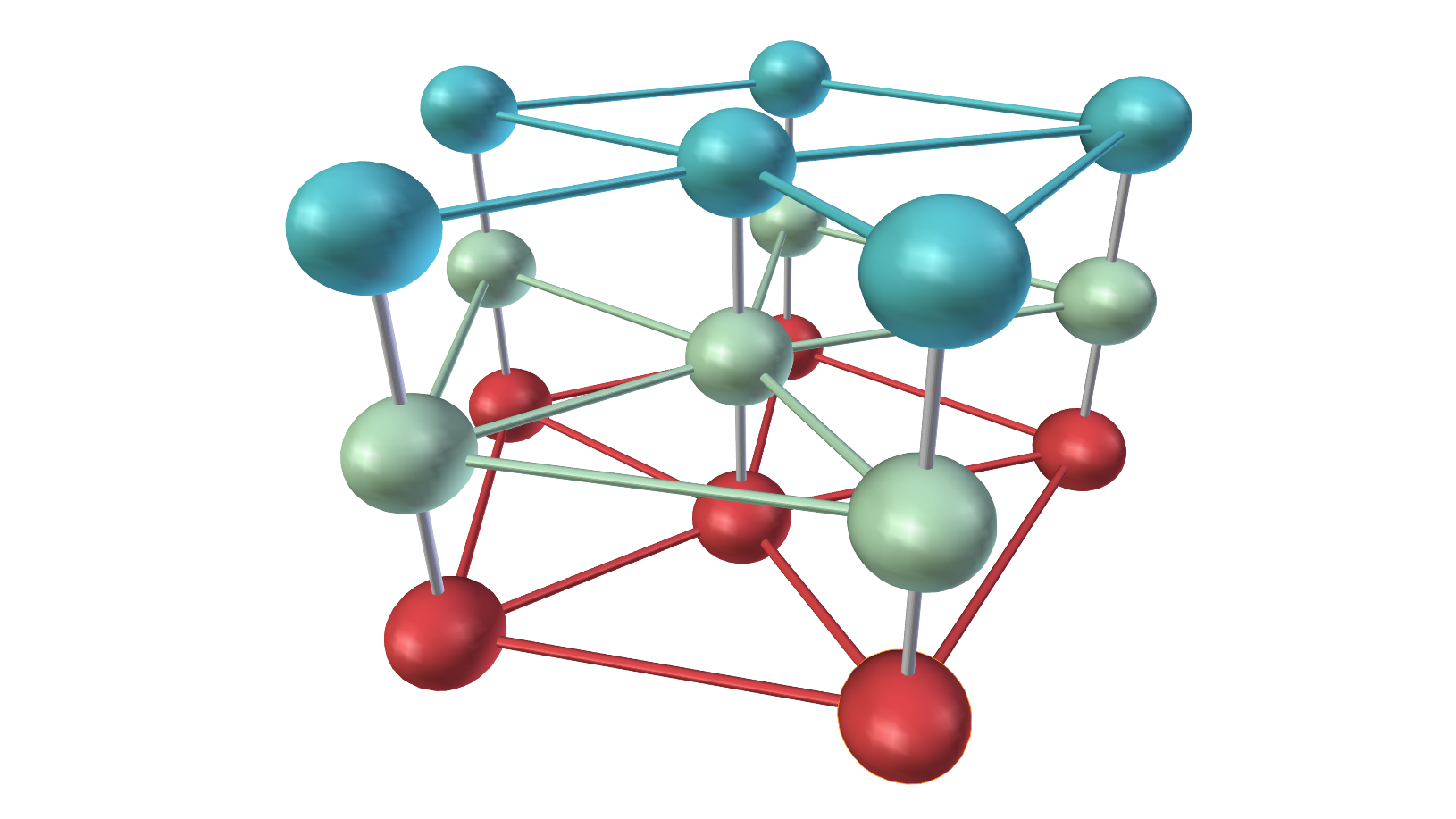}
\caption{(Color online) Example of multiplex network formed by three layers (blue, gray and red). The set of nodes is the same in all layers and the inter-layer links connect them in a one to one basis. For clarity, the inter-layer connection between the first and third layer has been omitted.
\label{fig:multiplex}}
\end{figure}

In a general multi-layer network \cite{kivela:2014} each node in one layer can be connected to any node in any other layer. Hence, the connectivity may be given by a tensor with four indices, $M_{ij}^{\alpha\beta}$, whose entries are 1 if nodes $i^\alpha$ --node $i$ in layer $\alpha$-- and $j^\beta$ are connected and zero otherwise. In multiplex networks \cite{gomez:2013, sole-ribalta:2013, dedomenico:2013, granell:2013, kouvaris:2015}, however, the set of nodes $i=1,\ldots,N$ is the same in all layers $\alpha=1,\ldots,L$, and the connectivity of each layer is given by a matrix $A^\alpha=\{ a^\alpha_{ij} \}$ (see Fig.\ref{fig:multiplex}). Furthermore, only connections among one node $i^\alpha$ and its counterpart in another layer $i^\beta$ are allowed. This inter-layer connectivity is the same for all nodes and is given by the inter-layer connectivity matrix $\Lambda^{\alpha\beta}$ \cite{sole-ribalta:2013}. Hence, multiplex networks have a connectivity defined by
\begin{equation}
\begin{split}
M_{ij}^{\alpha\alpha}=a^\alpha_{ij} \\
M_{ii}^{\alpha\beta}=\Lambda^{\alpha\beta} \\
M_{ij}^{\alpha\beta}=0
\end{split}
\end{equation}
with $i\neq j$ and $\alpha \neq \beta$.

Each site $i$ of a multiplex may be regarded as one structural entity, and the different layers $\alpha$ represent different interconnection structures between these entities. Structural entities (sites) may represent specific locations (cities), and the different connectivity of each transportation system or mobility pattern would define the different layers through which the agents move, while the inter-layer connections refer to the possibility to reach another transport (layer) from a specific location (for simplicity we will assume that $\Lambda^{\alpha\beta}=1$). Individuals changing from one layer to another can be represented as changing their strategy due to selection (due to prices, availability, or other competitive reasons) or mutation (random trial) in an evolutionary framework.

In order to describe the state of the entire multiplex system, we may define a set of vectors $\{\vec{n}_i\}$, one per site $i$, each one with $L$ (number of layers) components. From an evolutionary game theoretical point of view the components of such vectors represent the number of agents $n_i^\alpha$ in a given position $i^\alpha$ of the multiplex, and hence $\vec{n}_i$ is the population composition at $i$. Then, the evolution of the state of node $i^\alpha$ may be assumed as given by a functional $F_i^\alpha[\{\vec{n}_i(t)\}, a_{ij}^\alpha, \Lambda^{\alpha\beta}]$, which depends, respectively, on the state of the system, and on the intra- and inter-layer connectivities.

As the state of the system is instantaneously defined by a set of quantities $\vec{n}_i$ for each site $i$, we may 
write the dynamics as a set of coupled differential equations, one per component (layer),
\begin{equation} \label{eq:dep}
\frac{dn_i^\alpha(t)}{dt}= F_i^\alpha \left[n_i^\alpha(t),a_{ij}^\alpha n_j^\alpha(t), \Lambda^{\alpha\beta} n_i^\beta(t)\right],
\end{equation}
where the terms on which the functional depends are, respectively, the state of $i^\alpha$, the state of connected nodes in the same layer (intra-layer neighborhood of $i^\alpha$), and the state of equivalent nodes connected through inter-layer connections (inter-layer neighborhood of $i^\alpha$). 

In order to approach a replicator-equation like functional (see Eq.\eqref{eq:repeq}), we have to normalize $n_i^\alpha$ with respect to $N_i=\sum_\alpha n_i^\alpha$, the number of agents (population size) on site $i$, which is only site dependent. The normalized quantity is the fraction $x_i^\alpha=n_i^\alpha/N_i$, which must fulfill the restriction $\sum_\alpha x_i^\alpha=1$, and hence its derivative satisfies $\sum_\alpha \dot{x}_i^\alpha=0$. Note that, whenever the agents diffuse from $j^\beta$ to $i^\beta$ (or in the opposite direction), they are modifying the value of the fraction in $i^\alpha$ through the modification of the population size $N_i$. Hence, in order to account for the dynamics in a position $i^\alpha$ it is no longer enough to take into account the direct neighborhood of $i^\alpha$, given by its connectivity, but it is necessary to take into account the entire neighborhood of $i$, introducing an extra dependence. The functional in terms of fractions is thus
\begin{equation} \label{eq:depx}
\frac{dx_i^\alpha(t)}{dt}= F_i^\alpha \left[\vec{x_i}(t),a_{ij}^\alpha x_j^\alpha(t), \Lambda^{\alpha\beta} x_i^\beta(t), a_{ij}^\beta x_j^\beta(t)\right].
\end{equation}
where the extra dependencies are given by the first term between brackets, which now accounts for the entire state of $i$, and the last term, which accounts for the neighborhood of $i^\beta$. As we show in the following, such extra dependencies require a modification of the diffusion term, which will no longer be linear, but include a non-linear term.

\section{Replicator dynamics with diffusion in the multiplex}
\label{sec:diff}

In this section we derive a diffusion term in the multiplex compatible with the replicator dynamics, i.e. describing the evolution of the fractions of agents $x_i^\alpha$ at site $i$ with diffusive pattern given by layer $\alpha$, and keeping the conservation of the total number of agents in the multiplex, $\sum_i N_i=N$ where $N$ is constant, as well as the constraint $\sum_\alpha x_i^\alpha=1$. We will assume that diffusion and evolution are uncoupled, and hence the diffusion term can simply be added to the dynamics. 

As the diffusion of agents has been already studied, let us start trying to derive the equations for diffusion of the fractions from those for the number of agents, and discuss its use in game theoretical studies. If the agents are diffusing across a network structure with adjacency matrix of layer $\alpha$ given by the elements $a_{ij}^\alpha=1$ for connected nodes and $0$ otherwise, the transition probabilities determining the microscopic dynamics of diffusion of (numbers of) individuals $i$ given $j$ in layer $\alpha$ are
\begin{equation}\label{eq:transfalse}
\begin{tabular}{l}
$(T^{+\alpha}_{i|j})_\textnormal{diff} = D^\alpha a_{ij}^\alpha n^\alpha_j = D^\alpha a_{ij}^\alpha N_j x^\alpha_j$ \\
$(T^{-\alpha}_{i|j})_\textnormal{diff} = D^\alpha a_{ij}^\alpha n^\alpha_i = D^\alpha a_{ij}^\alpha N_i x^\alpha_i$
\end{tabular}
\end{equation}
where $D^\alpha$ is the diffusion coefficient of layer $\alpha$ and 
\begin{equation} \label{eq:transdiff}
T^{+\alpha}_{i|j} = T[n_i^\alpha \to n_i^\alpha +1 \, | \, n_j^\alpha \to n_j^\alpha - 1] ,
\end{equation}
is the transition rate of increase (decrease, by changing all signs) of the number of agents in $i^\alpha$ in one unit triggered by the movement of one agent in $j^\alpha$. Note that, as the process of diffusion impilies a redistribution of agents, the aggregated number of agents $N_i+N_j$ is preserved in each diffusive event among $i$ and $j$, and hence the constraint $(T^{+\alpha}_{i|j})_\textnormal{diff} = (T^{-\alpha}_{j|i})_\textnormal{diff}$ holds, which also ensures the global conservation of agents in the entire multiplex system by linking the processes $n_i^\alpha \to n_i^\alpha +1 $ and $n_j^\alpha \to n_j^\alpha - 1$. In the following, for simplicity, we will write the transition rates in a simplified form, explicitly stating the process to which it refers in the focal variables and the terms involved, but not the linked process; in this way, Eq.\eqref{eq:transdiff} would be $T[n_i^\alpha \to n_i^\alpha +1 \, | \, n_j^\alpha]$

The microscopic dynamics can be connected with the macroscopic behavior of the system by expanding a Fokker-Planck equation and truncating high order terms (this happens naturally for $N\gg1$ when working with fractions, see appendix~\ref{app:a} for the one dimensional derivation), which results in the Langevin equation,
  \begin{equation} \label{eq:landyn2D}
   \dot{\eta}_i^\alpha=e + \xi s 
  \end{equation}
where $\xi$ is uncorrelated Gaussian noise, $\eta$ is either the number of individuals or the fraction of individuals, and the drift and diffusion terms (do not confuse the latter with the diffusive dynamics studied in this section) are respectively
  \begin{equation} \label{eq:langevin2D}
   \begin{split}
   e=\sum_j (T^{+\alpha}_{i|j}-T^{-\alpha}_{i|j}), \\ s=\sqrt[]{\frac{\sum_j(T^{+\alpha}_{i|j}+T^{-\alpha}_{i|j})}{N_i}}.
   \end{split}
  \end{equation}
where the transition probabilities have to be written in terms of numbers or fractions depending on the choice in Eq.\eqref{eq:landyn2D}.
The first term in Eq.\eqref{eq:langevin2D} (drift) accounts for the deterministic behavior of the system and the latter (diffusion term) for stochastic effects. Note that the stochastic effects disappear in the thermodynamic limit $N_i\to\infty$, or whenever $N_i\gg \sum_j(T^{+\alpha}_{i|j}+T^{-\alpha}_{i|j})$, i.e. when the transition rates are small compared to the population size. 

Whenever we introduce the transition probabilities Eq.\eqref{eq:transfalse} into the drift term in Eq.\eqref{eq:langevin2D} expressed for numbers of individuals, the deterministic diffusive dynamics for the number of agents are
\begin{equation}\label{eq:diffn}
(\dot{n}^\alpha_i)_{\textnormal{diff}}=D^\alpha \sum_j a^\alpha_{ij}(n_j^\alpha-n_i^\alpha)=-D^\alpha \sum_j L^\alpha_{ij} n^\alpha_j
\end{equation}
well known equation for the diffusion of particles on a network, where the tensor $L^\alpha=\{L^\alpha_{ij}\}=\{\delta_{ij} k^\alpha_i - a^\alpha_{ij}\}$ is the graph Laplacian of the corresponding layer $\alpha$ (with $\delta^{\alpha\beta}$ the Kronecker's delta) and $k^\alpha_i$ is the degree of site $i$ in layer $\alpha$.

\begin{figure*}[!ht]
\begin{tabular}{lll} 
 &  (a) \phantom{En un lugar de la mancha} $\rho_{ij}=\frac{N_j}{N_i}$ & (b)\phantom{En un lugar de la mancha}$\rho_{ij}=1$ \\
 \begin{turn}{90} \phantom{En un lugar} Eq.\eqref{eq:diffx} (Full eq.). \end{turn} &  
 \begin{turn}{90} \phantom{En un lu} Fraction of individuals, $x_i^\gamma$. \end{turn} 
 \includegraphics[width=0.5\linewidth]{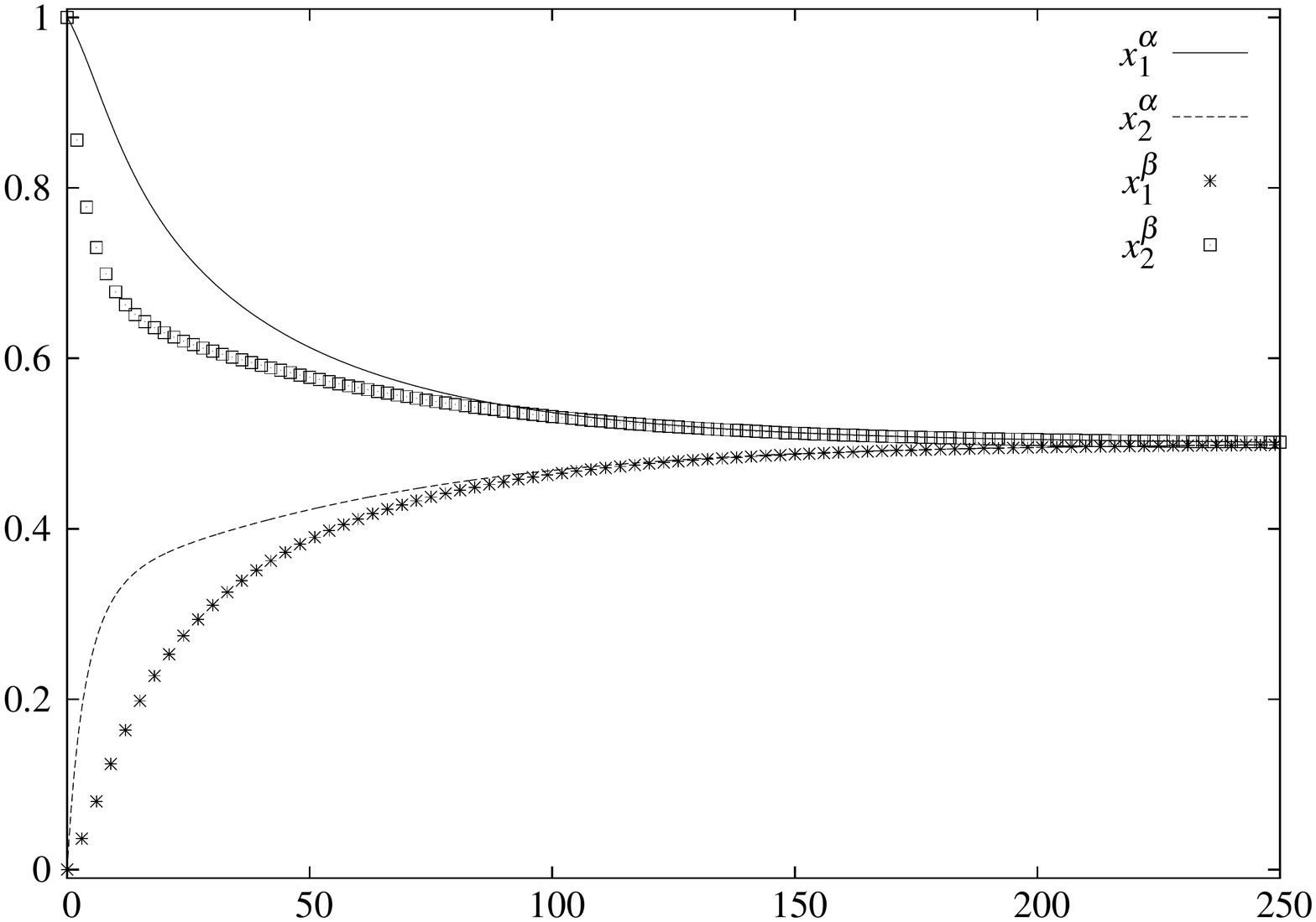} &  
 \includegraphics[width=0.5\linewidth]{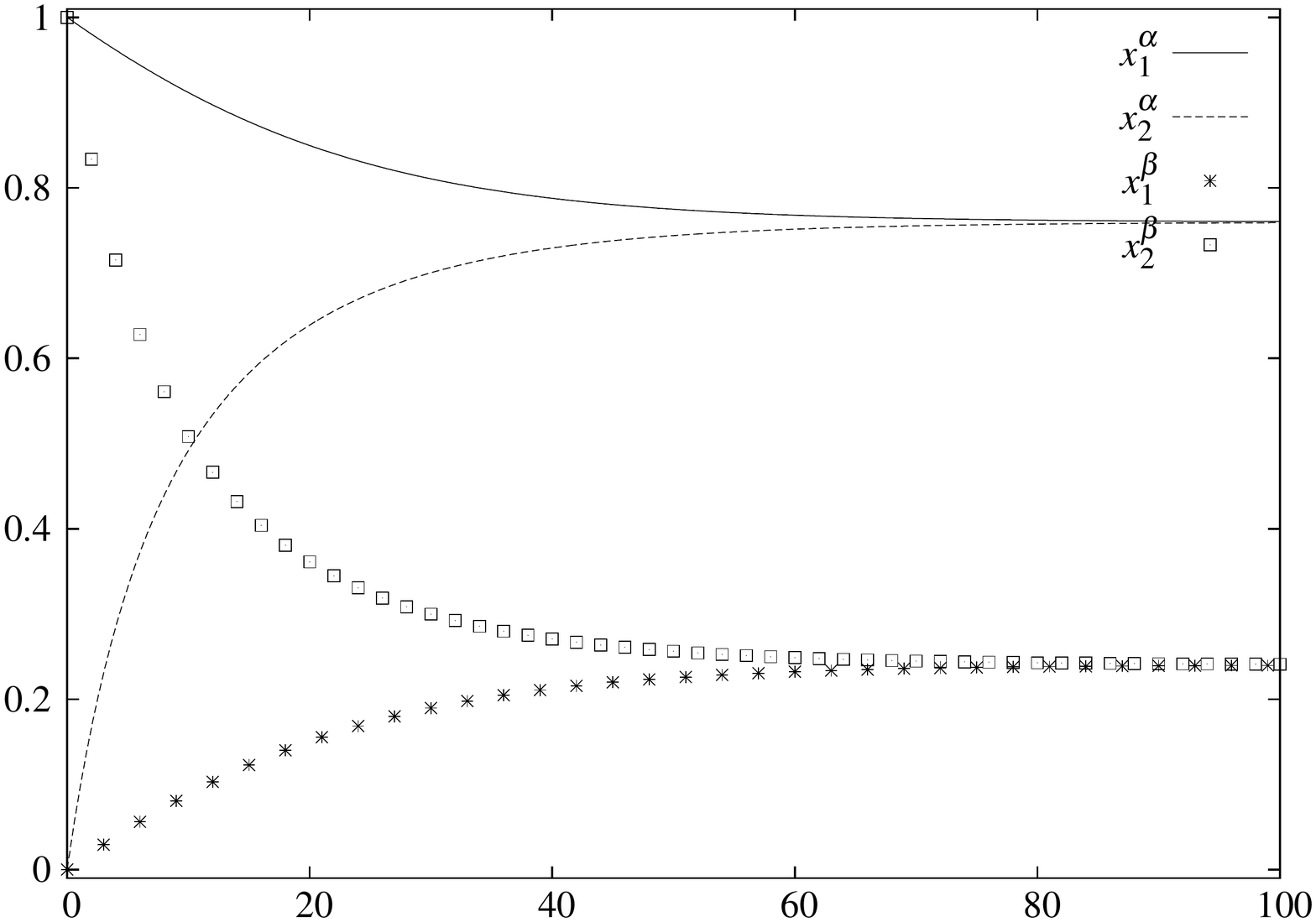} \\
 & (c) & (d) \\ 
 \begin{turn}{90} \phantom{En un lugar} Eq.\eqref{eq:difffalse} (linear). \end{turn} & 
 \begin{turn}{90} \phantom{En un lu} Fraction of individuals, $x_i^\gamma$. \end{turn}
 \includegraphics[width=0.5\linewidth]{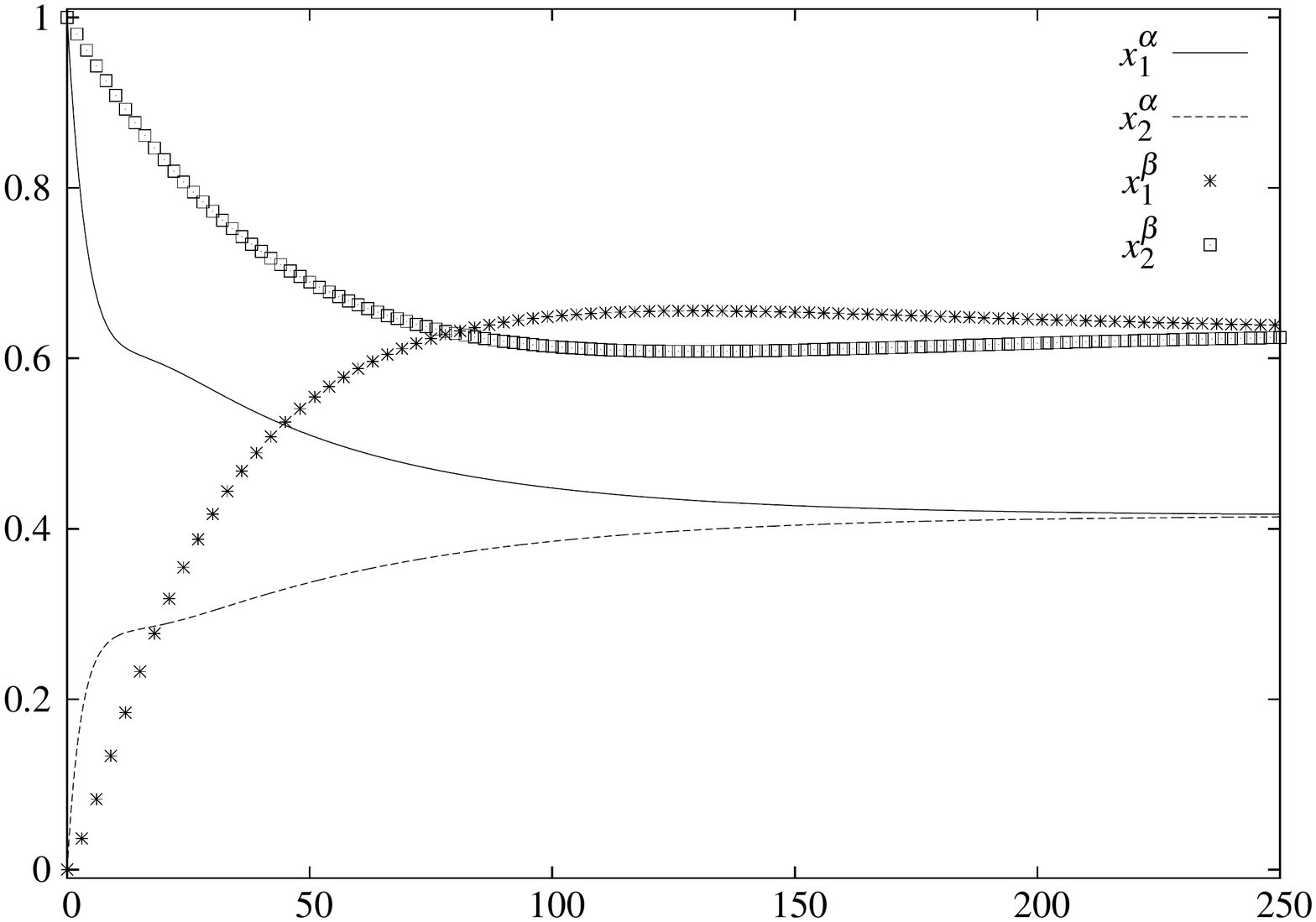}  &  
 \includegraphics[width=0.5\linewidth]{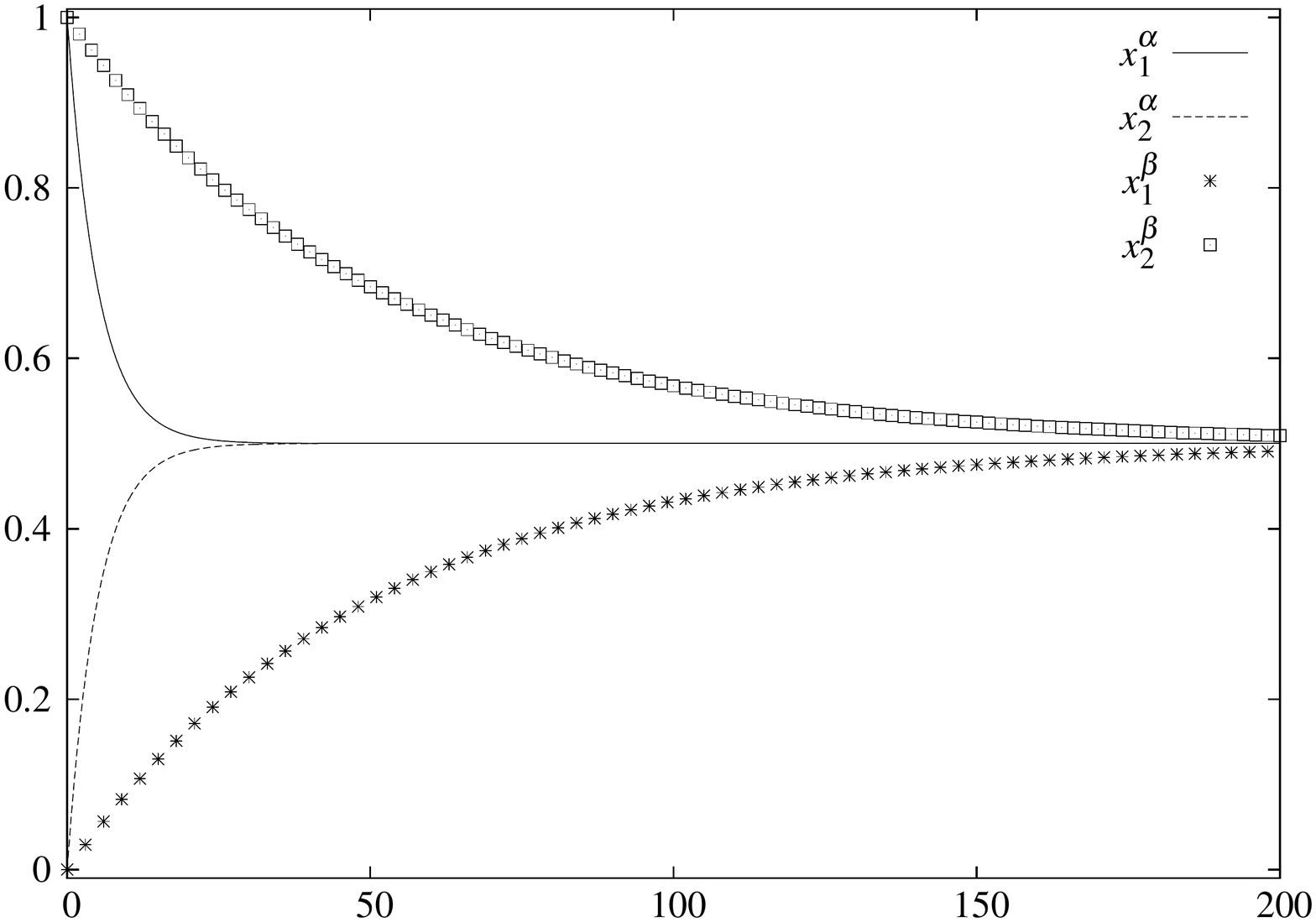} \\
 & \phantom{En un lugar de la mancha de } Time, $t$. & \phantom{En un lugar de la mancha de } Time, $t$.
\end{tabular}
\caption{Intra-layer diffusive dynamics of a system consisting of two layers with diffusion coefficients $D^\alpha=10^{-1}$ and $D^\beta=10^{-2}$, and two sites, $i=1$ and $j=2$ . The initial configuration is with $N$ individuals in nodes $i^\alpha$ and $j^\beta$ which are allowed to diffuse within its own layer (note that $N$ may take any value in the situation depicted, provided that it is the same in both layers, as the fractional character of $\rho$ makes it vanish); no inter-layer process is acting on the system. In order to fulfil the constraint $\sum_\alpha x_i^\alpha=1$, stars and solid line, and squares and dashed line, have to add up to one. (a,b) dynamics of Eq.\eqref{eq:diffx}, with (a) $\rho_{ij}$ calculated exactly analytically and (b) setting $\rho_{ij}=1$; (b,c) dynamics of Eq.\eqref{eq:difffalse} with (c) $\rho_{ij}$ calculated exactly analytically and (d) $\rho_{ij}=1$;  Cases (c) and (d) do not keep the normalization of the fractions. The dynamics given in (a) and (b) maintain the restriction $\sum_\gamma x_i^\gamma=1$ due to the quadratic term in Eq.\eqref{eq:diffx}. However, while case (a) describes accurately the expected dynamics, (b) shows an asymmetry in the final state, which is not expected due to pure intra-layer diffusion, but the result of an induced evolutionary pressure due to the different diffusive velocities and the restriction $\rho_{ij}=1$.
\label{fig:dif}}
\end{figure*}

In order to derive the equation for the diffusion of fractions instead of numbers, it is important to note that, if we differentiate the fraction $x^\alpha_i=n^\alpha_i/N$, we get
\begin{equation} \label{eq:repdiff}
\dot{x}^\alpha_i = \frac{\dot{n}^\alpha_i}{N_i}-x^\alpha_i\frac{\dot{N}_i}{N_i} 
\end{equation}
and hence a non-linear term that depends on $x^\alpha_i$ appears. Since we are trying to find the exact description of the diffusive process which is compatible with the replicator equation for mobile agents in a multiplex, we can follow this approach: first, construct a compatible macroscopic equation and then, infer the transition probabilities that give rise to it. In order to do this we may introduce Eq.\eqref{eq:diffn} into Eq.\eqref{eq:repdiff} (note that the latter is a replicator equation of the form of Eq.\eqref{eq:dx}), and use $\sum_\beta n_i^\beta=N_i$, obtaining a term of the form,
\begin{equation}\label{eq:diffx}
(\dot{x}^\alpha_i)_{\textnormal{diff}}=-D^\alpha \sum_j \rho_{ij} L^\alpha_{ij} x^\alpha_j + x_i^\alpha \sum_\beta \sum_j D^\beta \rho_{ij} L^\beta_{ij} x_j^\beta,
\end{equation}
where $\rho_{ij}=N_j/N_i$ is a population size dependence between neighboring sites. As it can be observed, in addition to a linear term, the first one, a second term appears. This term ensures that the normalization of the fractions is the proper one, i.e. fractions always add up to one, as shown in Fig.\ref{fig:dif}(a),(b) for a system of two layers and two nodes in each layer. Note that, as previously foreseen in section~\ref{sec:fractions}, this term implies a dependence of the dynamics at $i^\alpha$ on positions $j^\beta$ to which it is not directly connected, but represents the neighborhood of $i^\beta$ nodes.

The diffusion term \eqref{eq:diffx} may be rewritten in a more compact form
\begin{equation}\label{eq:diffxx}
(\dot{x}^\alpha_i)_{\textnormal{diff}}= - \sum_\beta \sum_j D^\beta x_j^\beta \rho_{ij} (\delta^{\alpha\beta} - x_i^\alpha) (k_{i}^\beta \delta_{ij} - a_{ij}^\beta)
\end{equation}
Note that, if one extracts the node degree from the second parenthesis --the network Laplacian term--, then it becomes $k_i^\beta(\delta_{ij}-a_{ij}^\beta/k_i^\beta)$, which is similar to the first parenthesis term, $\rho_{ij} (\delta^{\alpha\beta} - x_i^\alpha)$, but with constant values instead of variables. In this way the latter term may be interpreted as a population composition dependent Laplacian, which accounts for the instantaneous heterogeneity in population sizes and fractions of individuals in the network. 

Now, it is possible to infer the transition probabilities of the microscopic process from the emergent macroscopic dynamics (Eq.\eqref{eq:diffxx}), which may be used in Markovian analysis \cite{requejo:2012c}. The macroscopic deterministic dynamics emerge from the drift term in Eq.\eqref{eq:langevin2D}, which by extrapolation to two dimensions results in
\begin{equation} \label{eq:dyn3}
\dot{x}_i^\alpha=\sum_\beta \sum_j (T^{+\alpha|\beta}_{i|j}-T^{-\alpha|\beta}_{i|j}).
\end{equation}
where 
\begin{equation} \label{eq:trans3}
T^{+\alpha|\beta}_{i|j} = T\left[x_i^\alpha \to x_i^\alpha + \Delta x_i^\alpha \, | \, x_j^\beta \right] ,
\end{equation}
is the transition rate of increase (decrease, by changing all signs) of the fraction of agents in $i^\alpha$ given the fraction of agents in $j^\beta$ (see appendix~\ref{app:c}). Note that $x_j^\beta$ may also vary in the same process, but such variation is specified in its related transition rate, and not explicitly written here for simplicity.
 
Then, we can compare Eq.\eqref{eq:diffxx} and Eq.\eqref{eq:dyn3} (see appendix~\ref{app:c} for the derivation), and infer the bi-dimensional transition probabilities describing the variation of the fractions of individuals $x_i^\alpha$ which give rise to Eq.\eqref{eq:diffxx}, resulting in
\begin{equation} \label{eq:trans2d}
\begin{tabular}{ll}
$T^{+\alpha|\alpha}_{i|i} = 0$ &
$T^{-\alpha|\alpha}_{i|i} = D^\alpha k_i^\alpha x_i^\alpha (1-x_i^\alpha)$ \\
$T^{+\alpha|\alpha}_{i|j} = D^\alpha \rho_{ij} a_{ij}^\alpha x_j^\alpha (1-x_i^\alpha)$ &
$T^{-\alpha|\alpha}_{i|j} = 0$ \\
$T^{+\alpha|\beta}_{i|i} = D^\beta k_i^\beta x_i^\alpha x_i^\beta$ &
$T^{-\alpha|\beta}_{i|i} = 0$ \\
$T^{+\alpha|\beta}_{i|j} = 0$ &
$T^{-\alpha|\beta}_{i|j} = D^\beta \rho_{ij} a_{ij}^\beta x_i^\alpha x_j^\beta$ \\
\end{tabular}
\end{equation} 
where we have omitted the subscript 'diff' for simplicity. 

Note that the transition probabilities depend on both the layers ($\alpha$) and the sites ($i$), and that the symmetry which is present for mutation transition probabilities, which makes them keep local population sizes constant (see appendix~\ref{app:b}), is broken in general settings, $T^{-\alpha|\beta} \neq T^{+\beta|\alpha}$. Such asymmetry is however expected, as diffusion implicitly needs variable local population sizes (although the total number of particles in the multiplex is conserved), and assuming constant population sizes locally may have unexpected effects, as shown in the next section.

The transition probabilities in Eq.\eqref{eq:trans2d}, together with those that give rise to the replicator equation with mutations (appendices \ref{app:a},\ref{app:b}) complete the microscopic description of the evolutionary process for agents in a multiplex. Let us finally write down the deterministic replicator dynamics (see appendix~\ref{app:a}) with additive mutations (see appendix~\ref{app:b}) and diffusion in a multiplex:
\begin{equation} \label{eq:fulldyn}
\begin{split}
\dot{x}^\alpha_i= x^\alpha_i (f^\alpha_i-\bar{f_i}) + \sum_\beta (x_i^\beta q_i^{\beta\alpha} - x_i^\alpha q_i^{\alpha\beta}) \\
- \sum_\beta \sum_j D^\beta x_j^\beta \rho_{ij} (\delta^{\alpha\beta} - x_i^\alpha) (k_{s}^\beta \delta_{ij} - a_{ij}^\beta)
\end{split}
\end{equation}
where $f^\alpha_i$ is the fitness of individuals in position $i^\alpha$, $\bar{f_i}$ is the mean fitness of individuals in $i$ across layers and $q_i^{\alpha\beta}$ is the mutation rate which accounts for transitions of agents between $\alpha$ and $\beta$ layers in site $i$. For the case of mutations coupled to reproduction substitute the first two terms in the previous equation by the so called replicator-mutator equation (Eq.\eqref{eq:repmut}).

Remarkably, the non-linear effects due to diffusion are all contained in the $x_i^\alpha$ in the first parenthesis of the latter term in Eq.\eqref{eq:fulldyn}. There are however some situations in which the non-linearity disappears, and in which the analytic calculations can be simplified. Let us discuss them, as well as their implications, with special emphasis on the appearance of hidden selective pressures.

\section{Recovering linear diffusion and simplifying the analytics.}
\label{sec:linear}

Some situations allow us to recover linear diffusion, as well as to simplify the analytic calculations. Here, we analyze three of such scenarios. The first one refers to a situation in which population sizes are all forced to be constant across sites, and hence the second term in Eq.\eqref{eq:repdiff} disappears. Remarkably, in this situation a hidden selective pressure appears favoring the increase of fast diffusing strategies. The second scenario refers to slow population change, situation which approximates the constant population sizes scenario. The last subsection explores the situation in which the population ratio can be expressed as some analytic function. In that case, time scales separation allows us to write some formulas which simplify the calculations. 

\subsection{Constant population ratios (or sizes) induce hidden selective pressure.}

Let us first study the case in which some mechanism makes $\rho_{ij}=N_j/N_i \to c$, where $c$ is a constant; for simplicity, we will assume $c=1$, which includes the usual assumption in evolutionary game theoretical studies of constant population sizes \cite{helbing:1996, helbing:1998, traulsen:2005, szabo:2007, bladon:2010, traulsen:2012}). This case is shown in Fig.\ref{fig:dif}(b), where the restriction on the addition of fractions of individuals to one during the dynamical evolution of the system is fulfilled. However, the final state is not a symmetric one in which there are $N/2$ individuals in each node, as it happens without the restriction on the quotient of population sizes. Why does this happen?

The case in which $\rho_{ij}$ is forced to be equal to a certain value in Eq.\eqref{eq:diffx} may happen only if the conservation of particles due to the diffusive process does no longer hold. In such case, which may be due to a fast (compared to diffusion) and neutral evolutionary process acting on the system, the difference in diffusive velocities transforms into different "diffusive pressures", which induce different selective pressures while the system decays to the equilibrium. 

The different selective pressures are induced by the fact that fast diffusing individuals are increasing their frequency in a site compared to slower diffusing ones, and then the increased fraction is fixated by the neutral selective process, which only renormalizes the population size without altering the proportions. This is similar to a Wright-Fisher process, in which the population reproduces during the reproduction period according to their fitness and then, keeping the proportions of individuals, the population size is renormalized to its initial value. In our case, however, the variation in the fractions is due to the diffusion of the agents and the continuous renormalization of the population size is due to the neutral evolutionary process, thus favoring the fast diffusive strategy, as shown in Fig.\ref{fig:dif}(b).

\subsection{Slow population size change and quasi-neutral selection.}

The equation describing the evolution of the fraction of particles present at each point may in principle be simplified whenever, starting from a situation near the equilibrium, the variation of the number of agents is so slow that it can be neglected, $\dot{N_i}/N_i \to 0$. This is equivalent to assuming that $\dot{x}^\alpha_i \approx \dot{n}^\alpha_i/N_i $, which results in
\begin{equation}\label{eq:difffalse}
(\dot{x}^\alpha_i)_{\textnormal{diff}}=D^\alpha \sum_j a^\alpha_{ij}\left(\frac{N_j}{N_i} x_j^\alpha-x_i^\alpha\right) = -D^\alpha \sum_j \rho_{ij} L^\alpha_{ij} x^\alpha_j
\end{equation}
However, Eq.\eqref{eq:difffalse} does not generally keep the proper normalization for the fractions $\sum_\beta x_i^\beta = 1$, as it can be easily proven with a simple example. Let us assume a multiplex network formed by identical networks with identical diffusion coefficients (for two-dimensional spatial networks, this equals the Fisher-Kolmogorov reaction-diffusion scenario for gene wave-front propagation \cite{fisher:1937,kolmogorov:1937,fort:2002}). In this case, which is analogous to a mono-layer network, there is no dependence of the adjacency matrix terms and diffusion coefficients on the layer index $\alpha$, which now serves only to identify the different strategies present in each node. Hence, by summing Eq.\eqref{eq:difffalse} over layers (or strategies) and noting that $\sum_\beta x_i^\beta= 1$, we obtain the condition $\sum_\beta \dot{x}_i^\beta = (D/N_i) \sum_j a_{ij}(N_j -N_i) = (D/N_i) \sum_j a_{ij} N_j - k_i N_i$, which is only equal to zero, i.e. satisfies the dynamical constraint, whenever 
\begin{equation} \label{eq:const}
N_i=\frac{\sum_j a_{ij} N_j}{k_i} \,\, \text{or} \,\, N_i=N_j.
\end{equation}
These two restrictions are equivalent to requiring that $\sum_j L_{ij} N_j=0$, which could be implemented or engineered in the system, but it is not a priori expected to happen as a self-organizing feature. Furthermore, simulations using a system with two nodes and two layers confirm that the normalization is not fulfilled using Eq.\eqref{eq:difffalse}, as shown in Figs.\eqref{fig:dif}(c),(d). Hence, the construction of a microscopic model that describes the macroscopic diffusive dynamics of fractions of individuals cannot be done by simply rewriting the transition probabilities in Eq.\eqref{eq:transfalse} in terms of fractions.

The case of slow population change is equivalent to assuming that the second term in Eq.\eqref{eq:diffx} vanishes
\begin{equation}\label{eq:constn}
\sum_\beta \sum_j D^\beta \rho_{ij} L^\beta_{ij} x_j^\beta \to 0
\end{equation}
and is hence only slightly influencing the dynamics described by the first term. In such case, the diffusion term approaches Eq.\eqref{eq:difffalse}. Note however that, if we assume that the term above is strictly zero, we recover the case of strictly constant population sizes, and hence hidden selective pressures may appear, as explained in the previous subsection.

From an evolutionary perspective, small perturbations introduced in the \emph{neutral selection limit} satisfy the conditions leading to the slow population size change approximation whenever diffusion is slow. If we define the quasi-neutral selection limit as represented by $f^\alpha_i \to 0$, then the population size at each site varies slowly, given that $\dot{N_i} = N_i \bar{f}_i \to 0$ and $D^\gamma \to 0$. The quasi-neutral selection limit is important for two reasons: first, it allows for analytic calculations to be carried out in a similar way to the weak selection limit \cite{nowak:2004a, traulsen:2007} and, second, this limit approaches the neutral theory of molecular evolution proposed by Kimura \cite{kimura:1968}. However, care should be taken when approaching this limit, as explained above.

\subsection{Time-scales separation.}

Let us finish exploring the situations in which $\rho_{ij}$ can be written as a function of the fractions of individuals (or their fitness, which are determined by such fractions once the payoff matrix is known) and time due to time-scales separation.

Whenever we are able to write the population size dependence of the diffusion term as
\begin{equation}
\rho_{ij} = \rho_{ij}({x_k^\gamma},t).
\end{equation}
the entire replicator dynamics in the multiplex, including replication and deaths, mutations and diffusion, can be written as a function of the state of the system, given by the fractions of individuals in each node, and of the structural terms (diffusion coefficients and Laplacian of the multiplex).

In the cases in which it is not possible to write an explicit dependence for $\rho_{ij}$ as above, there are at least two situations in which time scales separation allows for approximations that take such form. The first one is whenever diffusion is slow and most of the population size change is due to replication and death. In this case it is easy to prove that the differential equation
\begin{equation}\label{eq:rho}
\dot{\rho}_{ij} \approx (\bar{f}_j-\bar{f}_i) \rho_{ij}
\end{equation}
governs the evolution of population sizes. Note that the solution of this equation is an exponential integral,
\begin{equation}\label{eq:memory}
\rho_{ij} (t)= \rho_{ij}^0 e^{\int_0^t (\bar{f}_j-\bar{f}_i) dt'},
\end{equation}
which implies that the system has memory. More precisely, the entire history of the difference of mean population fitness differences, which depends on the population compositions, is influencing the present state. Hence, Eq.\eqref{eq:memory} is a memory kernel of the past states of the system. 

As the memory kernel has an exponential form, although the entire history is contained in it, the influence of past states decays very fast with time compared to present states. This can be easily proved noting that, given two time lapses beginning at $t=0$ and ending at $t_1$ and $t_2>t_1$, the memory kernels are $\rho_{ij} (t_1)=\rho_{ij}^0 e^{\int_0^{t_1} (\bar{f}_j-\bar{f}_i) dt'}$ and $\rho_{ij} (t_2)=\rho_{ij}^0 e^{\int_0^{t_1} (\bar{f}_j-\bar{f}_i) dt'}e^{\int_{t_1}^{t_2} (\bar{f}_j-\bar{f}_i) dt'}$. The quotient between them is $\rho_{ij} (t_2)/\rho_{ij} (t_1)=e^{\int_{t_1}^{t_2} (\bar{f}_j-\bar{f}_i) dt'}$, which is a memory kernel of the time lapse between $t_1$ and $t_2$ and does not take into account the time lapse between 0 and $t_1$. Hence, we may always make $t_2=t_1+dt$ to express the quotient as an instantaneous integral of the fitness differences, which allows for computation of the dynamics without keeping track of the entire history of the system.

The opposite limit to slow diffusion is the fast diffusion limit: In this case we may assume that, after a short transient, the population size variations fulfill $\dot{N}_i/\dot{N}_j = \kappa_{ij}(t) \ne {0,\infty}$; then the approximation 
\begin{equation}
\rho_{ij} = (\bar{f}_i/\bar{f}_j) \kappa_{ij}(t)
\end{equation}
can be used. Furthermore, in the fast diffusion limit we may assume that the population size variations due to replication and death may be negligible compared to diffusion, as in the previous subsection; in such case the conditions in Eq.\eqref{eq:const} --equal populations across all sites or a local mean field describing the population sizes-- may be prone to happen, although again, this may introduce hidden selective pressures.

\section{Discussion and conclusions}
\label{sec:conclusions}

We have presented a complete description of the diffusive process in terms of fractions of individuals which is consistent with the replicator dynamics. Such term can be added to the replicator equation in order to make a full description of a system of diffusing and evolving agents in a multiplex. We have also found the transition probabilities that describe the microscopic dynamics from which the macroscopic behavior emerges. As we have shown, due to the multiplex structure it is necessary to include in the diffusion term a dependence on the quotient of population sizes of neighboring sites, as well as to add an extra diffusive term which is non-linear; this happens in order to keep the restriction imposed by the addition of fractions to one, as well as to account for the conservation of particles due to its diffusion. 

The extra non-linear diffusive term in Eq.\eqref{eq:diffx} takes into account the state of the vicinity of the focal node and its equivalent nodes in all layers (extended neighborhood), and not only its direct neighbors (directly connected nodes). This is relevant for the calculation of the evolution of the fractions of individuals at each site, and contrasts with the usual diffusion term, which only takes into account the state of directly connected nodes within each layer.

We have finally explored the recovery of the linear scenario, finding that when population sizes are constant across sites (due to environmental saturation, for instance), the non-linearity disappears. However, in this case the diffusion process induces an extra evolutionary pressure acting while the system reaches the equilibrium. This has been argued to happen because, even if the evolutionary process does not alter the fractions of individuals at each site, and only acts so as to maintain the population size constant, the faster diffusing strategy increases its proportion in the neighboring sites faster than the slower strategy in the focal site, and hence the renormalization process favors its increase. This suggests that, depending on the network architecture, the induced evolutionary selective pressures may work so as to create extra gradients of selection acting while the system is out of the equilibrium, which may induce, depending on the multiplex architecture, an unexpectedly complex phenomenology.

The authors acknowledge discussions about the ideas expressed in the paper with N.E.Kouvaris and J.Camacho, and the commentaries of two anonymous referees.
This work was supported by the LASAGNE (Contract No.318132) and MULTIPLEX (Contract No.317532) EU projects.
The authors acknowledges financial support from Generalitat de Catalunya (2014SGR608) and Spanish MINECO (FIS2012-38266).

\appendix
\section{Microscopic local derivation of the replicator equation.}\label{app:a}
\numberwithin{equation}{section}
\patchcmd{\theequation}{.}{}{}{}
The replicator equation in a well-mixed population (no network structure, Eq.(\ref{eq:repeq})) can be derived --for large population sizes accepting smooth derivatives-- by differentiating the fraction of individuals $x^\alpha$,
\begin{equation}\label{eq:dx}
\frac{d(x^\alpha)_\textnormal{rep}}{dt}=\frac{d(n^\alpha/N)_\textnormal{rep}}{dt}=x^\alpha\left( \frac{\dot{n}^\alpha}{n^\alpha}-\frac{\dot{N}}{N} \right)
\end{equation}
and assuming that the fitness of the individuals corresponds to the instantaneous per-capita growth rate of each strategy due to replication and deaths \cite{szabo:2007}, 
\begin{equation}\label{eq:fit}
f^\alpha({\bf x}) = \frac{\dot{n}^\alpha}{n^\alpha}, 
\end{equation}
where ${\bf x}=\{x^\alpha\}$ is the state vector of the population, and the mean fitness of the aggregated population fulfills $\bar{f}=\sum_\alpha x^\alpha f^\alpha = \dot{N}/N$. If the population size is constant, $\dot{N}=0$, 
then the per-capita growth rate is proportional to the difference between trait and mean population fitness,
\begin{equation}\label{eq:fitNcte}
f^\alpha({\bf x})-\bar{f}({\bf x})= \frac{\dot{n}^\alpha}{n^\alpha}, 
\end{equation}
being hence the replicator equation applicable for constant and variable populations with slightly different fitness definitions. As it will be proven in the following, the replicator dynamics also emerge as the macroscopic description of some microscopic dynamics.

Let us start the derivation of the replicator equation by assuming that there is a microscopic process, which can be described by some transition probabilities between states, and that such states are well defined. As we will assume that individuals of different types diffuse through different network architectures, each of such networks being part of a multiplex structure, let us introduce now the related notation: As before, each agent type will be labeled by the superscript $\alpha$ of the layer to which it belongs (related with its strategy), and the subscript $i$ will refer to a site in such layer.

The probability for node $i^\alpha$ to be at time $t$ in a state with $n_i^\alpha$ individuals of $\alpha$ type will be denoted as $P(n_i^\alpha,t)$, and the probability of increasing or decreasing such number of individuals by one individual will be $T^+(n_i^\alpha)$ and $T^-(n_i^\alpha)$ respectively, where
\begin{equation}
T^+(n_i^\alpha)=T[n_i^\alpha \to n_i^\alpha +1]
\end{equation}
(similarly for $T^-$ with a sign change). With this notation, it is possible to write the master equation
 \begin{equation}
  \begin{split}
    P(n_i^\alpha,t+1)-P(n_i^\alpha,t)= \phantom{lalecheenvinagrelalecheenvina} \\ P(n_i^\alpha-1,t-1) T^+(n_i^\alpha-1) 
     + P(n_i^\alpha+1,t-1) T^-(n_i^\alpha+1) \\
    - P(n_i^\alpha,t-1) [T^-(n_i^\alpha) + T^+(n_i^\alpha)]
   \end{split}
  \end{equation}
which describes the evolution of $n_i^\alpha$. Now, let us assume that the total number of agents in site $i$ is $N_i=\sum_\alpha n_i^\alpha \gg 1$. We do not require it to be infinite, but just large enough, so that we can make a continuous approach without neglecting finite size fluctuations. In this case it is possible to define the re-scaled variables $x_i^\alpha=n_i^\alpha/N_i$, $\tau=t/N_i$ and $\rho(x_i^\alpha,\tau)=N_i P(n_i^\alpha,t)$. For simplicity, let us assume that there are only two strategies present in the population, and the constraint $\sum_\alpha x_i^\alpha=1$; we can then expand the master equation in a one dimensional Taylor expansion for $N_i \gg 1$ (the derivation is similar for the three-- \cite{bladon:2010} and n--strategies cases \cite{traulsen:2012} by using a multivariate Taylor expansion), giving rise to 
  \begin{equation}
   \begin{split}
   \frac{d}{dt}\rho(x^\alpha_i,t)=-\frac{d}{dx^\alpha_i}[e(x^\alpha_i)\rho(x^\alpha_i,t)] \\ +\frac{1}{2}\frac{d^2}{(dx^\alpha_i)^2}[s^2(x^\alpha_i)\rho(x^\alpha_i,t)].
   \end{split}
  \end{equation}

As the previous equation has the form of a Fokker-Planck equation, it is possible to transform it into the Langevin equation
  \begin{equation} \label{eq:landyn}
   \dot{x}_i^\alpha=e(x_i^\alpha) + s(x_i^\alpha) \xi
  \end{equation}
where $\xi$ is uncorrelated Gaussian noise and the drift and diffusion terms are
  \begin{equation} \label{eq:langevin}
   \begin{split}
   e(x_i^\alpha)=T^+(x_i^\alpha)-T^-(x_i^\alpha), \\ s(x_i^\alpha)=\sqrt[]{\frac{T^+(x_i^\alpha)+T^-(x_i^\alpha)}{N_i}}
   \end{split}
  \end{equation}
respectively accounting for the deterministic behavior and the stochastic effects. 

Note that, whenever the transition probabilities can be written as 
\begin{equation} \label{eq:transRD}
T^+(x_i^\alpha)=x_i^\alpha R(x_i^\alpha), \phantom{m} T^-(x_i^\alpha)=x_i^\alpha D(x_i^\alpha),
\end{equation}
the drift term $e(x_i^\alpha)$ looks like a replicator equation. This is indeed the only term acting in the thermodynamic limit $N \to \infty$, and whenever the size of large populations does not increase or decrease too fast, $N_i \gg x_i^\alpha (R(x_i^\alpha)+ D(x_i^\alpha))$. In this cases the Langevin equation simplifies to
\begin{equation} \label{eq:lanrep}
\dot{x}_i^\alpha=x_i^\alpha\cdot(R(x_i^\alpha)-D(x_i^\alpha))
\end{equation}
and the terms $R(x_i^\alpha)$ and $D(x_i^\alpha)$ act as the replication and death components of the fitness difference $f_i^\alpha-\bar{f}_i$ (compare Eqs.\eqref{eq:lanrep} and \eqref{eq:repeq}). Hence, any 
process in which the transition probabilities can be factorized as shown in Eq.\eqref{eq:transRD}, can be written as a replicator like system with fitness obtained as the solution of the equation
\begin{equation}
f_i^\alpha-\bar{f}_i=R(x_i^\alpha)-D(x_i^\alpha).
\end{equation}
This property has a particularly useful value: by using the latter equation and decomposing fitness in payoff components, it is possible to combine several processes, as virus spread (linear models) and cultural reproduction (frequency dependence, usually non-linear), into a unified evolutionary framework, as shown in \cite{requejo:2015}.

\section{Microscopic local updating rules and mutations}\label{app:b}
\numberwithin{equation}{section}
\patchcmd{\theequation}{.}{}{}{}
Whenever there is an arbitrary number of strategies in the population, the transition rates in Eq.\eqref{eq:transRD} can be decomposed in additive terms in a way in which each term relates to the contribution of each of the strategies. In this way Eq.\eqref{eq:landyn} can be generalized to
\begin{equation} \label{eq:dyn2}
\dot{x}_i^\alpha=\sum_\beta (T^{+\alpha|\beta}_i-T^{-\alpha|\beta}_i),
\end{equation}
where $T^{+\alpha|\beta}_i$ is the transition rate of increase of the fractions of individuals in $i,\alpha$  corresponding to the increase in one unit due to the action of agents in $\beta$,
\begin{equation}
T^{+\alpha|\beta}_i=T\left[n_i^\alpha \to n_i^\alpha +1 \, | \, n_i^\beta\right] ,
\end{equation}
and decreases them by one unit for the minus sign (note that this may imply a variation in the local population size $N_i$). The exact choice of the transition rates depends on the microscopic dynamics. These transition rates may be written as \cite{bladon:2010},
\begin{equation} \label{eq:trans2}
T^{+\alpha|\beta}_i = \sum_\gamma x_i^\gamma x_i^\beta g_i^{+\gamma|\beta} q_i^{\gamma\alpha}\, , \,\,\, T^{-\alpha|\beta}=T^{+\beta|\alpha}
\end{equation}
whenever the transitions depend on a big number of random interactions between agents in the same site (local well-mixing assumption) in which one individual replaces another (defined by the second condition) and there are mutations between $\gamma$ and $\alpha$ individuals at a rate $q_i^{\gamma\alpha}$. Factor $g^{\pm\gamma|\beta}$ contains the information about how the microscopic updating rule acts: it states the exact mechanism by which one strategy increases or decreases due to the action of another.

Two probabilistic microscopic dynamics are usually investigated which give rise to the replicator dynamics (assuming $q_i^{\gamma\alpha}=\delta^{\gamma\alpha}$). The first one is the modified Moran process, where one randomly chosen individual is assumed to die, and another individual, chosen according to a probability proportional to its fitness, reproduces. This process is defined by $g_i^{+\alpha|\beta} = f_i^\alpha/\bar{f_i}$ and $g_i^{-\alpha|\beta} = g_i^{+\beta|\alpha}$. The second process is proportional imitation, where one individual compares its strategy to another one, both chosen at random in the mean field limit, and changes its strategy with a probability increasing linearly with the difference of the payoffs between them. This process is defined by $g_i^{+\alpha|\beta} = (1/2)(1+(f_i^\alpha-f_i^\beta)/\Delta f_{s,\textnormal{max}})$ and  $g_i^{-\alpha|\beta} = g_i^{+\beta|\alpha}$, where $\Delta f_{s,\textnormal{max}}$ is the maximum fitness difference and keeps the proper normalization. In both cases the symmetry condition $g_i^{-\alpha|\beta} = g_i^{+\beta|\alpha}$ ensures that the evolutionary process maintains a constant population, and the dynamics result in the replicator equation, up to a multiplicative factor which relates to the temporal scale of the dynamics.

Whenever mutations happen as a strategy change at any point during the lifetime of the individuals, they can be introduced as additive terms \cite{helbing:1998} of the form 
\begin{equation} \label{eq:tranmut}
(T^{+\alpha|\beta})_\textnormal{mut}=x^\beta q^{\beta\alpha} \, , \,\,\, (T^{-\alpha|\beta})_\textnormal{mut}=(T^{+\beta|\alpha})_\textnormal{mut}
\end{equation}
to the transition rates in Eq.\eqref{eq:trans2}, where the coupled mutations may be eliminated by setting $q_i^{\alpha\beta}=\delta^{\alpha\beta}$ (with $\delta^{\alpha\beta}$ the Kronecker's delta). The introduction of the additive term in the transition rates gives rise to the extra additive term in the replicator dynamics 
\begin{equation}\label{eq:mut}
(\dot{x}_i^\alpha)_\textnormal{mut}=\sum_\beta (x_i^\beta q_i^{\beta\alpha} - x_i^\alpha q_i^{\alpha\beta}).
\end{equation}
This term describes the effect of random mutations or equivalently of random exploration of strategies \cite{traulsen:2009} in the evolutionary process. For the case of equal symmetric mutations, i.e. $q_i^{\alpha\beta}=q_i^{\beta\alpha}=\mu$, between all strategies this term simplifies to 
\begin{equation}\label{eq:mutmu}
(\dot{x}_i^\alpha)_\textnormal{mut}=\mu(1-L x_i^\alpha)
\end{equation}
where $L$ is the number of strategies \cite{arenas:2011}.

Let us finally recall that, when mutations are coupled to the reproductive dynamics as in Eq.\eqref{eq:trans2}, then Eq.\eqref{eq:dyn2} gives rise to the replicator-mutator equation,
\begin{equation}\label{eq:repmut}
(\dot{x}_i^\alpha)_\textnormal{rep,mut}=\sum_\beta x_i^\beta f_i^\beta q_i^{\beta\alpha} - x_i^\alpha \bar{f_i}.
\end{equation}
The diffusion term in Eq.\eqref{eq:diffxx} could also be added to this equation to represent situations where only newborns mutate.

\section{Derivation of the transition probabilities for the non-linear diffusion term} \label{app:c}

The microscopic dynamics that give rise to the non-linear equation describing the dynamics of diffusion for fractions of individuals (Eq.\eqref{eq:diffxx}) is defined by the transition probabilities determining the increase or decrease of the number of individuals, as shown in Eq.\eqref{eq:dyn3}. In order to infer such transition probabilities, which complete the bottom up description, we can start expanding Eq.\eqref{eq:diffxx} as
\begin{equation}
\begin{tabular}{l}
$(\dot{x}^\alpha_i)_{\textnormal{diff}}=$ \\
$- \sum_\beta \sum_j D^\beta x_j^\beta \rho_{ij} (\delta^{\alpha\beta} k_{i}^\beta \delta_{ij} - \delta^{\alpha\beta} a_{ij}^\beta - k_{i}^\beta x_i^\alpha  \delta_{ij} + a_{ij}^\beta x_i^\alpha)$
\end{tabular}
\end{equation}

Then, we can split the double sum on the four contributions corresponding to the terms arising from $i^\alpha$, $j^\alpha$, $i^\beta$ and $j^\beta$ (assuming that $\alpha \neq \beta$ and $i \neq j$), obtaining
\begin{equation}
\begin{tabular}{l}
$(\dot{x}^\alpha_i)_{\textnormal{diff}}= 
 -D^\alpha k_i^\alpha x_i^\alpha (1-x_i^\alpha) + \sum_{j\neq i} D^\alpha a_{ij}^\alpha \rho_{ij} x_j^\alpha (1-x_i^\alpha)$ \\
$+ \sum_{\beta \neq \alpha} D^\beta k_i^\beta x_i^\alpha x_i^\beta
- \sum_{j\neq i} \sum_{\beta \neq \alpha} D^\beta a_{ij}^\beta \rho_{ij} x_i^\alpha x_j^\beta$ 
\end{tabular}
\end{equation}
respectively.

By performing the same kind of split in Eq.~\eqref{eq:dyn3} we find
\begin{equation}
\begin{tabular}{l}
$ \dot{x}_i^\alpha=
 T^{+\alpha|\alpha}_{i|i}-T^{-\alpha|\alpha}_{i|i} + \sum_{j \neq i} (T^{+\alpha|\alpha}_{i|j}-T^{-\alpha|\alpha}_{i|j}) $ \\
$ + \sum_{\beta \neq \alpha} (T^{+\alpha|\beta}_{i|i}-T^{-\alpha|\beta}_{i|i}) 
+ \sum_{j \neq i} \sum_{\beta \neq \alpha} (T^{+\alpha|\beta}_{i|j}-T^{-\alpha|\beta}_{i|j})$
\end{tabular}
\end{equation}

Then, since the terms relate to the influence of different nodes $k^\gamma$ on the focal node $i^\alpha$, we can compare term by term both equations above, finding that each term in the first equation corresponds to the substraction of a pair of terms in the second. 

In order to finally find the right transition terms, we need to take into account the physical constraints in the system. Let us analyze them one by one. 

First, we can focus on the first term on the above equations. This term relates to the influence of $i^\alpha$ on its own dynamics. Since the agents in $i^\alpha$ are diffusing away from that position, the influence is necessarily negative, and hence $T^{+\alpha|\alpha}_{i|i}=0$, and 
\begin{equation}
T^{-\alpha|\alpha}_{i|i}=T\left[ \frac{n_i^\alpha}{N_i} \to \frac{n_i^\alpha-1}{N_i-1} \,|\, x_i^\alpha \right]=D^\alpha k_i^\alpha x_i^\alpha (1-x_i^\alpha)
\end{equation}
where $| x_i^\alpha$ denotes a dependence, not implying that such term is constant (indeed, it varies in the process). Note also that, whenever $x_i^\alpha=0$, the transition rate is zero, as expected due to the absence of agents diffusing away. Furthermore, if $x_i^\alpha=1$, the transition rate is again zero, as expected due to the fact that agents diffusing away decrease the number of agents, but leave the fraction unchanged.

Now, let us focus on the influence of $j^\alpha$ on the dynamics of $i^\alpha$ ($j\neq i$). Since agents diffusing away from $j^\alpha$ are incresing the number of agents in $i^\alpha$, the influence is necessarily possitive, and the transition rates are $T^{-\alpha|\alpha}_{i|j}=0$ and
\begin{equation}
T^{+\alpha|\alpha}_{i|j}=T\left[ \frac{n_i^\alpha}{N_i} \to \frac{n_i^\alpha+1}{N_i+1} \,|\, x_j^\alpha \right]=D^\alpha a_{ij}^\alpha \rho_{ij} x_j^\alpha (1-x_i^\alpha)
\end{equation}
The limits, as before, can be easily proven to behave in the correct way.

Then, for the influence of $i^\beta$ on $i^\alpha$ ($\beta \neq \alpha$), we have to note that agents diffusing away from $i^\beta$ decrease the denominator of the fraction $x_i^\alpha=n_i^\alpha/\sum_\gamma n_i^\gamma$, since they are decreasing $n_i^\beta$, and hence the fraction $x_i^\alpha$ increases. Therefor, the transition rates are $T^{-\alpha|\beta}_{i|i}=0$ and
\begin{equation}
T^{+\alpha|\beta}_{i|i}=T\left[ \frac{n_i^\alpha}{N_i} \to \frac{n_i^\alpha}{N_i-1} \,|\, x_i^\beta \right]=D^\beta k_i^\beta x_i^\alpha x_i^\beta
\end{equation}

Finally, individuals diffusing away from $j^\beta$ ($j\neq i$ and $\beta \neq \alpha$) are incresing the number of individuals in $i^\beta$, and hence increasing the denominator in the fraction of $x_i^\alpha$ (opposite as in the previous case). The influence is then so as to decrease the fraction of individuals at $i^\alpha$ and hence $T^{+\alpha|\beta}_{i|j}=0$ and 
\begin{equation}
T^{-\alpha|\beta}_{i|j}=T\left[ \frac{n_i^\alpha}{N_i} \to \frac{n_i^\alpha}{N_i+1} \,|\, x_j^\beta \right]=D^\beta a_{ij}^\beta \rho_{ij} x_i^\alpha x_j^\beta
\end{equation}

From the transition probabilities above we can derive the diffusion term corresponding to the stochastic effects in the Langevin equation (extrapolating $s$ to two dimensions in Eq.\eqref{eq:langevin2D}), which is
\begin{equation} \label{eq:diffdrift}
\small
s=\sqrt{
\frac{
\sum_\beta \sum_j D^\beta x_j^\beta \rho_{ij} (x_i^\alpha + (1-2x_i^\alpha) \delta^{\alpha\beta}) (a_{ij}^\beta+k_i^\beta \delta_{ij})
}{N_i}
}
\end{equation}
This term can be added to Eq.\eqref{eq:diffxx} --multiplied by white Gaussian noise-- in order to account for the stochastic effects introduced by the diffusive process in situations in which the thermodynamic limit cannot be assumed, or population sizes vary fast. 


\end{document}